\documentclass[a4paper,fleqn,usenatbib,article]{mnras}

\usepackage{newtxtext,newtxmath}
\usepackage[T1]{fontenc}
\usepackage{ae,aecompl}

\usepackage{graphicx}	% Including figure files
\usepackage{amsmath}	% Advanced maths commands

\usepackage{amssymb}	% Extra maths symbols

\title[Dust changes in Sakurai's Object]{Dust changes in Sakurai's Object: new PAHs and SiC with coagulation of submicron-sized silicate dust into 10$\mu$m-sized melilite grains}

\author[Bowey]{
Janet E. Bowey$^{1}$\thanks{E-mail: boweyj@cardiff.ac.uk}
\\
$^1$ School of Physics and Astronomy, Cardiff University, Queens Buildings, The Parade, Cardiff CF24 3AA, UK.
}

\date{Accepted XXX. Received YYY; in original form ZZZ}

\pubyear{2020}

\begin{document}
\label{firstpage}
\pagerange{\pageref{firstpage}--\pageref{lastpage}}
\maketitle

\begin{abstract}

6--14~\micron\ Spitzer spectra obtained at 6 epochs between April 2005
and October 2008 are used to determine temporal changes in dust
features associated with Sakurai's Object (V4334 Sgr), a low mass
post-AGB star that has been forming dust in an eruptive event since
1996. The obscured carbon-rich photosphere is surrounded by a
40-milliarcsec torus and 32 arcsec PN. An initially rapid mid-infrared flux
decrease stalled after 21 April 2008. Optically-thin emission due to
nanometre-sized SiC grains reached a minimum in October 2007,
increased rapidly between 21--30 April 2008 and more slowly to October
2008. 6.3-\micron\ absorption due to PAHs increased
throughout. 20~\micron-sized SiC grains might have contributed to the
6--7~\micron\ absorption after May 2007.  Mass estimates based on the
optically-thick emission agree with those in the absorption features
if the large SiC grains formed before May 1999 and PAHs formed in
April--June 1999. Estimated masses of PAH and large-SiC grains in
October 2008, were $3\times 10^{-9}$~M$_{\sun}$ and
$10^{-8}$~M$_{\sun}$, respectively. Some of the submicron-sized
silicates responsible for a weak 10~\micron\ absorption feature are
probably located within the PN because the optical depth decreased
between October 2007 and October 2008.  6.9~\micron\ absorption
assigned to $\sim$10~\micron-sized crystalline melilite silicates
increased between April 2005 and October 2008. Abundance and
spectroscopic constraints are satisfied if $\la $2.8 per cent of the
submicron-sized silicates coagulated to form melilites. This figure is
similar to the abundance of melilite-bearing calcium-aluminium-rich
inclusions in chondritic meteorites.
\end{abstract}


%%%%%%%%%%%%%%%%%%%%%%%%%%%%%%%%%%%%%%%%%%%%%%%%%%

%%%%%%%%%%%%%%%%% BODY OF PAPER %%%%%%%%%%%%%%%%%%

\section{Introduction}
\defcitealias{Evans2020}{Ev2020} Sakurai's Object (V4334 Sgr) is a low
mass post-AGB star that is undergoing a very late thermal pulse caused
by the ignition of a residual helium shell. It has formed substantial
quantities of dust during the 25 years since its discovery by Yukio
Sakurai in 1996 \citep{1996IAUC.6322....1N}. The source was later
identified (V$\sim$ 12.5~mag) in pre-discovery optical images from
January 1995; it brightened to 11.4~mag in the first 12 months and
showed no nova-like emission lines. In March 1996 the photosphere was
hydrogen-poor with over abundances of carbon and oxygen and centred in
a 32~\arcsec circular planetary nebula
\citep{1996ApJ...468L.111D}. Once the photosphere was obscured CN,
C$_2$ and CO were detected in its atmosphere \citep{Eyres1998}. Later
infrared observations revealed HCN and C$_2$H$_2$ with $^{12}$C to
$^{13}$C isotope ratios which are consistently lower than the Solar
System values;\citep[see][hereinafter~\citetalias{Evans2020}, for a
  detailed history]{Evans2020}. Like other `Born Again Giants' it lies
at the centre of an older faint planetary nebula (PN)
\citep{Pollacco1999}. \citet{Chesneau2009} obtained high
spatial-resolution 7.5--13.5~\micron\ observations in June 2007 and
deduced that the star was surrounded by a $30\times40$ milliarcsec
opaque dusty disc or torus inclined at 75~$\deg$ to the plane of the
sky and aligned with a low level of asymmetry seen in the old PN. The
source continues to evolve and the 2015--2019 expansion of the bipolar
nebula has recently been mapped in the near-infrared by
\citet{Hinkle:2020} and images show the existence of 760~K dust which
are likely to be a consequence of a near-IR brightening event observed
in 2008 \citep[e.g.][]{Hinkle2014}.

\begin{figure*}
  %landscape
  \includegraphics[bb=23 40 505 755,height=\linewidth,clip=,angle=-90]{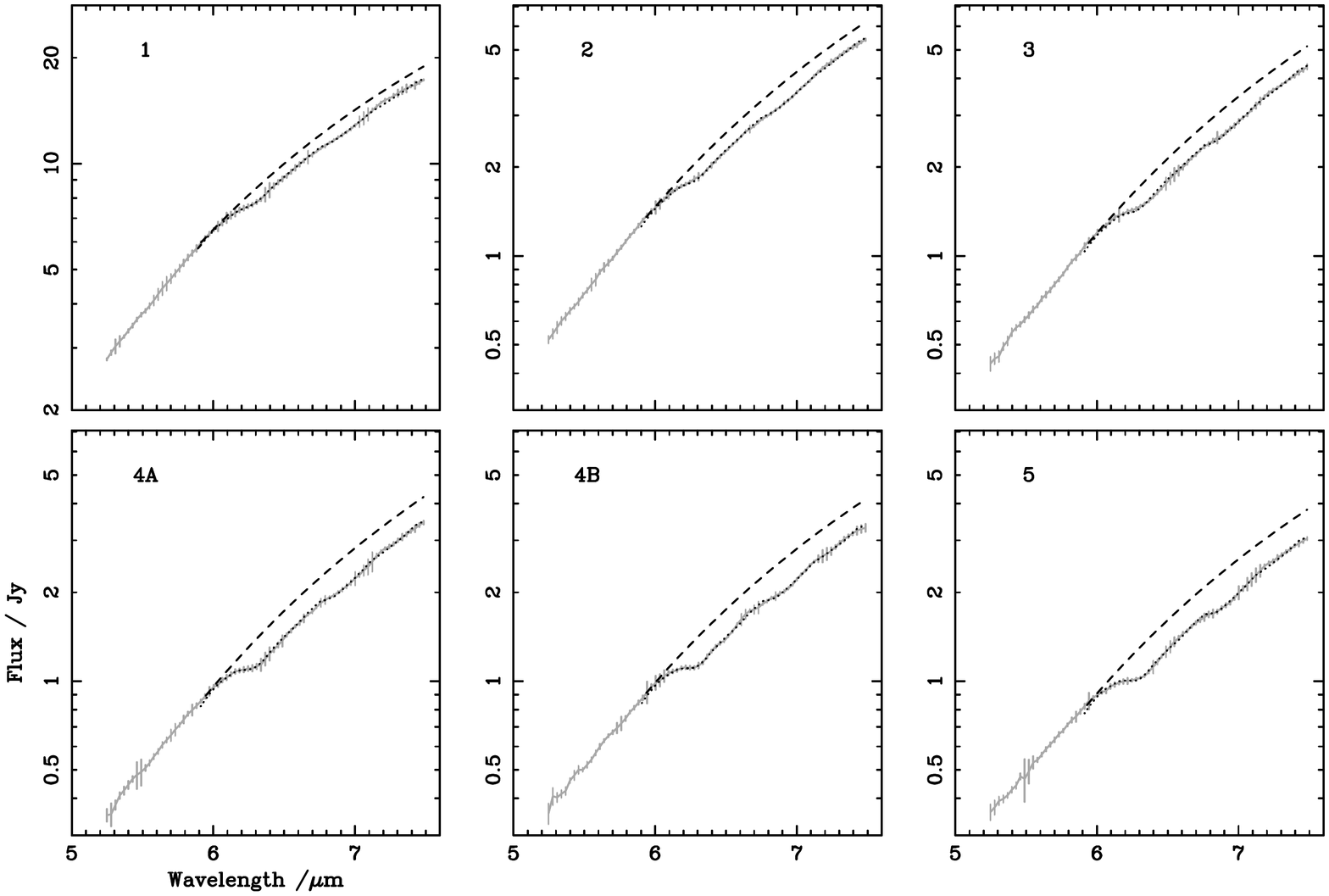}
\caption{5.2--7.5\micron\ spectra of Sakurai's object (grey error
  bars) with features fitted with a black body obscured by PAHs,
  melilite and bSiC (dotted; equation~\ref{eq1}) and derived continua
  (dashed). \label{fig:sl2}}
\end{figure*}

To date the dust mineralogy has not been studied in much detail; this
work is focussed on \citetalias{Evans2020}'s reporting of a weak, but
consistent, absorption feature with minima at 6.3~\micron\ and
6.9~\micron\ (Figure~\ref{fig:sl2}) in their low-resolution Spitzer
spectra of the source which they tentatively attributed to
hydrogenated amorphous carbon (HAC) formed in an early mass ejection
phase prior to 1997.  These features resemble part of similar
absorption bands in dense cold ($\la 100$~K) lines of sight through
young stellar objects (YSOs) and molecular clouds
\citep[e.g.][]{Keane2001, Boogert2011} which are normally associated
with a combination of ices, carbonaceous material, carbonates, or
occasionally with overtone bands of a crystalline silicate called
melilite \citep{BH2005}. Overtone bands are 1/100th the strength of
fundamental bands like the 10~\micron\ Si-O stretch and are therefore
seen only in samples where the fundamental bands are opaque due to
larger grain sizes or in micron to 100~\micron-thick films of
finely-ground powders. Since the grain density in YSO discs is
extremely high \citep{BH2005} did not make the connection with large
grain sizes or coagulated grains producing the features. In Sakurai's
Object the absence of evidence for H$_2$O ice at 3.0~\micron~and
6.0~\micron~in the currently oxygen-poor environment means that H$_2$O
and other ices and carbonates (which are not known to form in the
absence of water) can be ruled out as carriers of the 6.3~\micron~and
6.9~\micron~ bands. However, its complex history of mass-loss and
planetary nebula would suggest that carbonaceous materials, silicates
and other refractory components could contribute to the absorption
features.

The Spitzer observations used in this analysis are described in
Section~\ref{sec:obs}. Models of the 5.9--7.5~\micron~and
8.3--13.3~\micron~spectra are developed and the choice of laboratory
data explained in Sections~\ref{sec:sl2mod} and~\ref{sec:sl1mod}.  The
fits to each range are described in Sections~\ref{sec:sl2fits}
and~\ref{sec:8--13fit} and the observational results summarised in
Section~\ref{sec:csil}. A time-averaged dust profile for the
6--7.5~\micron-range is produced in Section~\ref{sec:wmean}. Dust
column mass and number densities are quoted in Section~\ref{sec:est}
together with estimates of the masses of carbonaceous dust formed.
The evolution of the dust during the Spitzer observations is discussed
in Section~\ref{sec:evo}. Conclusions are in
Section~\ref{sec:conclusion}.

\section{Observations}
\label{sec:obs}
\begin{figure}
\includegraphics[bb=0 125 266 326,width=\linewidth,clip=]{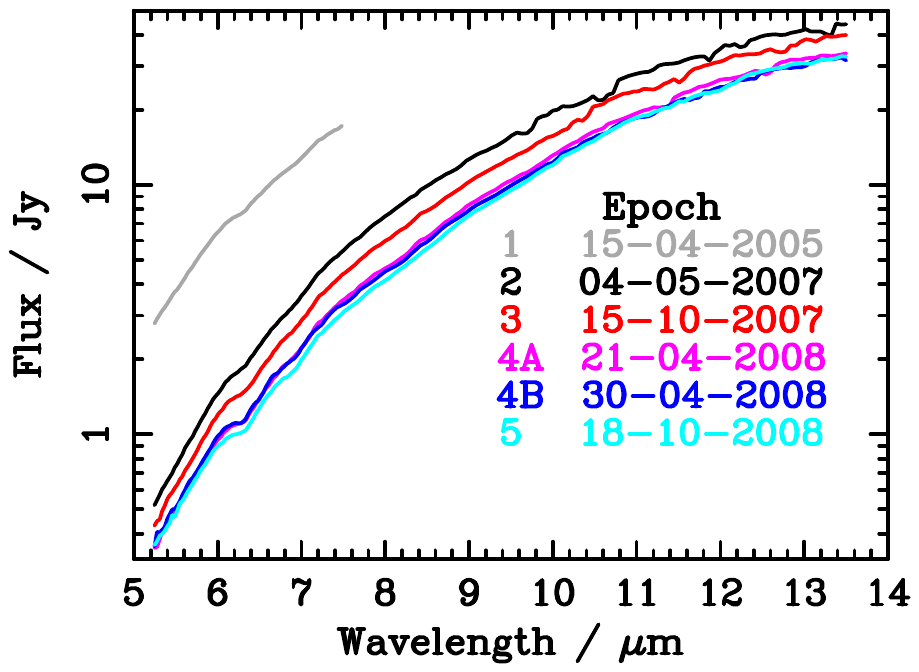}
\caption{Merged Spitzer SL2 (5--7.5~\micron), SL3
  (7.5--8.5~\micron) and SL1 (7.6--14.0~\micron)--spectra of
  Sakurai's Object for Epochs 2--5. SL2 only for Epoch 1; SL1 data were saturated.\label{fig:allflux}}
\end{figure}

\begin{table*}
\centering
\caption{Spitzer LRS Observations \citepalias{Evans2020}}
\label{tab:observations}
\begin{minipage}{\linewidth}
\begin{tabular}{llllll}
\hline
Epoch&MJD\footnote{Modified Julian Date (MJD) is used to identify the time of the observations. MJD is related to Julian Date (JD) by MJD = JD -- 2400000.5}
&Date &AOR &T/K&$F_{5.6}$/Jy\footnote{Weighted mean of 5.25--6.00~\micron\ flux.}\\ 
\hline
1&53475 &2005-04-15&10840320&284&$3.54\pm0.01$\\
2&54225 &2007-05-04&17740544&226&$0.921\pm0.003$\\
3&54388 &2007-10-15&17742336&217&$0.734\pm0.003$\\
4A&54577&2008-04-21&17742592&-- &$0.591\pm0.003$\\
4B&54586&2008-04-30&22273280&-- &$0.604\pm0.002$\\
5&54757 &2008-10-18&22272768&207&$0.571\pm0.002$\\
\hline
\end{tabular}
\end{minipage}
\end{table*}

Sakurai's Object was observed with the Low-Resolution Spectrometer
(LRS) \citep{Houck2004} on the Spitzer Space Telescope
\citep{Werner2004} on six occasions between 15 April 2005 and 18
October 2008 at intervals of 750, 163, 189, 9, and 171 days,
respectively designated Epochs~1, 2, 3, 4A , 4B and 5 in
Table~\ref{tab:observations} and Figure~\ref{fig:allflux}.  The PI was
A. Evans and the programmes were 3362, 30077 and 40061. The
5.6~\micron\ flux fell by a factor of 4 in the between April 2005 and
April 2007 (Epochs 1--2) and stabilised at 1/6th of the original flux
between April and October 2008 (Epochs 4--5).

The Spitzer observing modes and results from both high (HR) and low
resolution (LR) observations are described by \citep[][AOR~
  10840320]{Evans2006} and \citetalias{Evans2020}. Pipeline-reduced
and optimally-extracted spectra covering orders SL~1, ~2 and ~3 used
in this paper were obtained from the Combined Atlas of Sources with
Spitzer/IRS Spectra (CASSIS) \citep{Leb2011} in 2020; CASSIS used data
reduction pipeline S18.18.0. Fits were constrained with the combined
systematic and RMS errors (the most pessimistic uncertainties);
AOR~17742592 SL~1 observations were constrained only with RMS errors
because this part of the CASSIS spectrum did not include systematic
errors. The observations are shown with the fitted models in
Figures~\ref{fig:sl2} and \ref{fig:sl1fluxemiss}(a).

Due to the disjointed nature of both the laboratory and observational
data, I initially fitted 5.3-7.5~\micron\ SL~2 spectra and then
extended the analysis to the SL~1 (7.6--14.0\micron) and SL~3
(7.5--8.5~\micron) bands in a piecemeal fashion. SL~1 and SL~3
spectral segments were scaled by up to $\pm 5$ percent to match the
SL~2 fluxes and the weighted means of interleaved regions obtained
before fitting them.  SL~1 data were trimmed at 13.5~\micron\ before
analysis to exclude a spectral artefact known as the
14-\micron\ teardrop \citep[see][]{IRShandbook}. HR spectra cover the
the 10~\micron\ to 37~\micron\ range which excludes the PAH bands and
half the 10~\micron\ astronomical silicate feature modelled
here. Slopes of scaled 9.9--13.5~\micron\ HR spectra match those of the
LL spectra, but the HR data are otherwise beyond the scope of this work.

\section{Model for 5.9~to 7.5~\micron~SL~2 Spectra}
\label{sec:sl2mod}
In order to determine the absorption optical depth profile, the likely
dust components, and a continuum, the Spitzer SL~2 flux spectra were
fitted with an obscured black body model with up to three foreground
absorption components, where the flux, $F_\nu$ is given by:
\begin{equation}
F_\nu=c_0B_\nu(T)\exp{(-\sum^3_{i=1} c_i\uptau_i(\lambda))},
\label{eq1}
\end{equation}
where $B_\nu(T)$ is the Planck function, and $\uptau_i$ is the shape
of the $i^{th}$ absorption feature, normalised to unity at the tallest
peak in the wavelength range of interest, and $c_0$ and $c_1$ to $c_3$
are the fitted scale factors. $c_0$, was determined so that the
continuum was matched to the feature at 6.0\micron, i.e. the
foreground absorption was assumed zero at this wavelength. Absorption
components were selected on the basis of their ability to the peak and
width of observed absorption bands with a preference for fewer
components and an improvement in $\chi^2$ values if more components
were added. Absorption components were fitted simultaneously using the
down-hill simplex method of $\chi^2$-minimization. Laboratory data
used in fitting are listed in Table~\ref{tab:lab}, described in
Appendices~\ref{app:pah}, and ~\ref{app:mhs}, and the preferred
candidates are discussed below. Optical depth profiles were obtained
for each epoch by deriving a source continuum with the foreground
absorption set to zero, i.e.  $\uptau=\ln{(F_{o}/c_0B_\nu(T))}$, where
$F_{o}$ is the observed spectrum.

Due to the narrow wavelength ranges involved in fitting in comparison
to the breadth of the peak of the Planck function, the black body
temperature will drift to unrealistic values and other parameters will
not settle if it is unconstrained. Therefore temperature was selected
by trial and error to provide the lowest $\chi^2$. Values used were
consistent with those obtained by \citetalias{Evans2020}; where they
differ significantly (Epoch~5), this is likely to be due to the
$<8\mu$m spectra being insensitive to the cooler dust components. The
current model does not include a term for the interstellar reddening
(powerlaw extinction) because, the visual extinction is low
\citep[A$_V\sim 2$; ][]{Evans2002}, the photosphere obscured and the
temperature relatively unconstrained.  Fits to the flux spectra and
continua are in Figure~\ref{fig:sl2}. The derived optical depth
profiles of dust towards Sakurai's Object for the different Epochs and
the laboratory spectra used in fitting are compared in
Figure~\ref{fig:optplot}.

\begin{table*}
\centering
\caption{Laboratory data and band asignments of the C-rich (PAH, bSiC and nSiC) and O-rich (melilite and astrosilicate) dust components used in the models; mass absorption coefficients are given at the peak wavelengths marked in bold type; see the appendices for further information}
\label{tab:lab}
\begin{minipage}{0.75\linewidth}
\begin{tabular}{llllllll}
\hline
Sample&Bands&Approximate&Size\footnote{Representative grain length assuming approximately cubic geometry.}&$\rho$&$\kappa_{pk}$&$m_g$\footnote{Representative mass of single grain, assuming volume=(size)$^3$ }&Ref\footnote{Reference for spectrum (also see appendices): 1--\citealt{Carpentier2012}; 2--\citealt{Hof2009};  3--\citealt{BH2005}; 4--\citealt{Speck2005};5--\citealt{BA2002}}\\
&&Assignment&&gcm$^{-3}$&$10^2$cm$^2$g$^{-1}$&g&\\
\hline
PAH&{\bf 6.3}, 6.9&Arom. C=C&53$_{30}^{70}$~nm&$0.39^{0.2}_{1.8}$&$600_{130}^{1200}$\footnote{Best estimates. The superscripts and subscripts indicate the range depending on the effective mass
  density of the soot sample, the value of 1200 pertains to a grain of 30nm and bulk mass density; see Appendix~\ref{app:pah}}&$5.8\times 10^{-17}$&1\\
&7.3&stretch&&&&\\
&8.0&C--C defect&&&&\\
&11.3, 11.9&Arom. CH bend&&&\\
%\hline
bSiC&6.2, \bf{6.6}&Overtones&25~\micron&3.2&2.4&5.0$\times 10^{-8}$&2\\
&7.1, 7.6&&&&\\
nSiC&{\bf 12.3}&Si--C stretch&3~nm&3.2&$15000$&$8.6\times10^{-20}$&2, 4\\
\hline

melilite\footnote{The formula of this  melilite sample is
  (Ca$_{1.7}$Na$_{0.3}$)(Fe$_{0.2}$Mg$_{0.4}$Al$_{0.4}$)(Al$_{0.1}$Si$_{1.9}$)O$_7$);
  it is an Si-rich intermediate member of the {\aa}kermanite
  (Ca$_2$MgSiO$_7$) to gehlenite (Ca$_2$Al(AlSi)O$_7$) solid-solution
  series in which one of the Si atoms is replaced by Al.}
&6.1, {\bf 6.9}&Overtones&12~\micron&3.0&2.2&$5.2\times 10^{-9}$&3\\

astrosilicate\footnote{Represented by the line of sight to Cyg OB2 no.12; $\rho$ is an estimate given by the mean of forsterite and enstatite mass densities; $\kappa_{pk}$ from \citet{BA2002}.}&{\bf 9.8}&Si--O stretch&0.3~\micron&3.3&26&$8.9\times 10 ^{-14}$&\\
\hline
\end{tabular}
\end{minipage}
\end{table*}

\subsection{6.3\micron: Polycyclic Aromatic Hydrocarbons (PAHs)}

\begin{figure*}
  %landscape
  \includegraphics[bb=23 40 505 755,height=\linewidth,clip=,angle=-90]{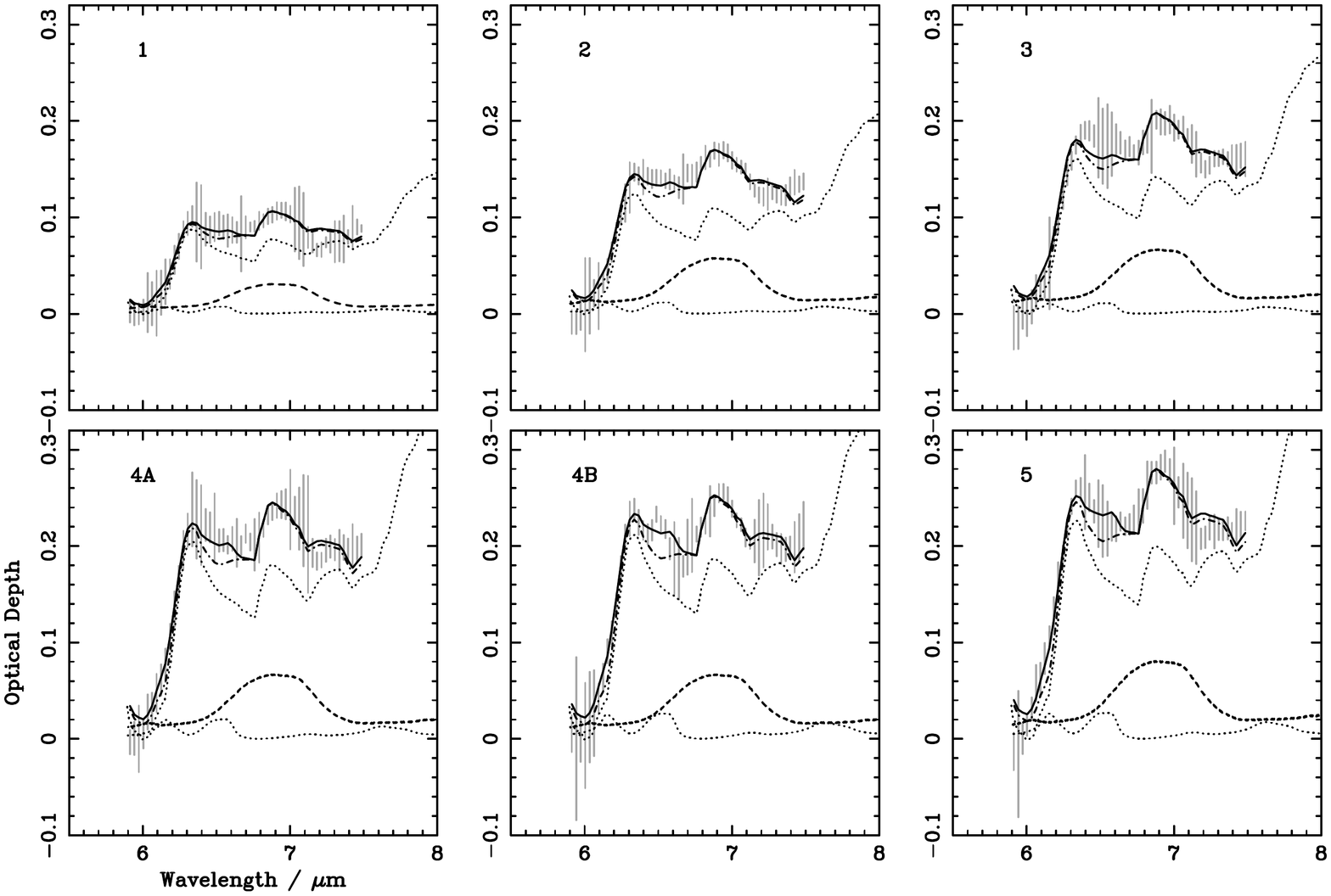}
\caption{6--7.5~\micron\ optical depth profiles of Sakurai's Object
  (error bars) with fits (solid); fitted PAH (dotted -top), bSiC (dotted) and melilite (dashed) components. Dot-dash curves indicate the effect of removing the bSiC component from the fits.\label{fig:optplot}}
\end{figure*}

The Hydrogenated Amorphous Carbon (HAC) compounds suggested by
\citetalias{Evans2020}, were not a good match to the features since
the proposed spectra \citep{Grishko2002} are dominated by a broad
merged 5.8~\micron~(attributed to carbonyl C$=$O) and 6.2~\micron~peak
which is not observed in Sakurai's Object (see
Appendix~\ref{app:HAC}). Good eye-ball single-component matches to the
$\sim $6.3~\micron~ absorption peak and the position of the $\sim
6.9$-\micron~peak were obtained with the spectrum of a PAH-dominated
soot sample created and characterised by \citet{Carpentier2012} (see
Figure~\ref{fig:optplot}).  However, these PAH bands are too
structured to match the strength and breadth of 6.9~\micron~absorption
band without an additional component. I have not found a carbonaceous
material with a sufficiently broad absoption feature at this
wavelength \citep[e.g.][presented 56 spectra of PAHs in this
  wavelength range]{Gavilan2017} and but given the complexity of dust
formation this possibility cannot be completely ruled out.

\subsection{Melilite as the carrier of the broad $6.9\mu$m band}

\subsubsection{Absence of ice and carbonate towards Sakurai's Object}
The similar YSO and molecular-cloud 6.9-\micron\ feature is frequently
associated with methanol ice or carbonates, but neither produce an
ideal match \citep[e.g.][]{Keane2001,Boogert2011, BH2005}. Of these
candidates methanol is preferred because its abundance in molecular
clouds and YSOs is 3--30\% of H$_2$O ice \citep[e.g.][]{Whittet2003}.
However, the low visual extinction and absence of a H$_2$O-ice
detection towards Sakurai's Object makes this identification
implausible in this environment. In addition methanol ice has a
narrower band, more similar to the similar to the PAH
6.9-\micron~band, than to the missing absorption component. The
absence of water in the environment would also preclude the existence
of carbonates which form in the presence of liquid water within
asteroid bodies or terrestrial environments \citep[e.g.][]{AB2005}.

\subsubsection{Oxygen-rich candidates: melilite and hibonite}
\label{sec:omel}
Even though Sakurai is currently C-rich the presence of the old PN
indicates that this may not always have been the case.  Dual dust
chemistries have been observed in PNe associated with Wolf-Rayet
stars with PAH emission below $\sim 15\mu$m and crystalline silicates
at longer wavelengths \citep{Cohen2002}; features from both C-rich and
O-rich materials are also seen in the oxygen-rich PN of NGC~6302 which
displays prominent PAH emission as well as emission from crystalline
silicates \citep{Molster2001} and \citep{Hofmeister2004} associated
far-infrared emission bands with melilite and hibonite. In the
near-infrared, \citet{BH2005} presented overtone spectra of $\sim 30$
refractory minerals present in meteorites and showed that the broad
6.9~\micron~ bands in YSOs are matched reasonably well by an overtone
feature seen in 12~\micron-thick samples of compressed crystalline
melilite powder, i.e. with an effective grain size of $\sim 12\mu$m,
or possibly $\sim~60~\mu$m-sized grains of disordered (metamict)
hibonite ($\sim$CaAl$_{12}$O$_{19}$) (see
Appendix~\ref{app:mhs}). These large grains have opaque fundamental
bands (e.g. Si--O stretches in melilite) and therefore indetectable at
the wavelengths where astronomical mineral searches are normally
conducted. Only the melilite feature is observed because melilites
have particularly strong overtone features which are three times as
strong those of other anhydrous crystalline silicates
(e.g. forsterite, pyroxenes, felspars) and ten times those of
forsterite and diopside glass \citep[see][]{BH2005}.

Melilites are an important component of calcium-aluminium-rich
inclusions (CAIs) in primitive meteorites.  CAIs are early
high-temperature condensates in solar nebular theories
\citep[e.g. review by][]{RubinMa2017}. CAI abundance in the most
primitive (CO3) chondrites is $\sim 1.5$ area per
cent\footnote{i.e. on average 1.5 per cent of the surface area of the
  sampled meteorite slices were comprised of CAIs; this number is very
  similar to the volume per cent. Some other chondrite types,
  especially CVs have more, with $\sim$ 3 area per cent on average
  \citep[e.g.][]{2008M&PS...43.1879H}.} with an average CAI size of
$98.4\pm 54.4$~\micron\ \citep{Zhang2020}. Melilite-rich inclusions
and spinel-pyroxene CAIs comprise 80-100 percent of the total CAIs for
each chondrite; \citeauthor{Zhang2020} argued that the spinel-pyroxene
CAIs in these and more processed meteorites are alteration products
derived from melilites and found that a subset of fluffy melilites
were composed of loosely aggregated $<15$\micron-sized melilite
grains.

Hibonite is less plausible because the fits are poorer, fewer than
4.8 per cent of CAIs in meteorites are hibonite (and grossite;
$\sim$CaAl$_4$O$_7$)-bearing \citep{Zhang2020}, with the metamict form
being relatively rare and other hibonites have weaker overtones which
do not occur at 6.9~$\mu$m, and hibonite bands are bands are 5-30
times weaker than the melilite feature \citep[see
  Appendix~\ref{app:mhs} and][]{BH2005}.

Since Sakurai's Object has had a complicated history of mass-loss and
is surrounded by a PN and an interstellar silicate absorption feature
was detected by \citet{Evans2002} and no other candidate has been
found I modelled the data with a melilite component
(Figure~\ref{fig:optplot}). Statistics of two component fits
($\chi^2_{\nu 2}$) are given in Table~\ref{tab:sl2fits} and the
sum of the PAH and melilite components of the fit to Epoch~5 indicated
by the dashed curve in Figure~\ref{fig:optplot}.

\subsection{A component due to 25\micron\ -sized Carbide Grains?}
\label{sec:bigsic}

Isotope measurements exist for thousands of 0.1--20~\micron-sized
meteoritic grains with non-solar isotope ratios of pre-solar origin
\citep[e.g.][]{Hoppe1994,Speck2005} suggesting that they somehow
travelled from C-stars to the Solar Nebula. However, SiC has not been
detected in the interstellar medium \citep[e.g][]{Whittet1990}. This
is probably a consequence of its high opacity, since grains have to be
nanometre-sized to produce reliable unsaturated bands
\citep[see][]{Hof2009}. However, $\la 25~\mu$m SiC grains, hereinafter
denoted bSiC (for ``big'' SiC grains), produce small peaks at 6.2~ and
6.6~\micron, whilst being opaque in their fundamental Si-C stretch at
12.3~\micron~\citep[][see Appendix~\ref{app:mhs} for its
  derivation]{Hof2009}, so I included them in three-component models.
Due to slight rounding of the 6.6~\micron\ peak, 25~\micron\ is likely
to be an upper size limit (see Appendix~\ref{app:mhs}).

\section{5.9 -- 7.5\micron\ Fits}
\label{sec:sl2fits}
\begin{table*}
\centering
\caption{Fits to 5.9--7.5~\micron~spectra of Sakurai's object: black
  body absorbed by PAHs, melilite, and bSiC. The 1 sigma confidence intervals
  $\sigma$, for $c_{pah}$ are 1--2 per cent; for $c_{mel}$ they are
  normally 4--5 per cent and 7 per cent in Epoch~1. $\uptau_{6.3}$ and $\uptau_{6.9}$ are the measured optical depths of the 6.3 and 6.9\micron\ peaks, respectively.
  $\chi^2_{\nu}$ and $\chi^2_{\nu2}$ denote three-component and two-component (PAH and melilite only) fits, respectively.}
\label{tab:sl2fits}
\begin{minipage}{\linewidth}
\begin{tabular}{llllllllll}
\hline
Ep. &T/K  &$c_{pah}$&${c_{mel}}$&$c_{SiC}$&$\sigma$&$\chi^2_{\nu}$&$\chi^2_{\nu2}$&$\uptau_{6.3}$&$\uptau_{6.9}$\\
\hline
1 &275  &  0.085&0.031&$<$0.007&59&0.90&0.91&0.10&0.11\\
2 &226  &0.13&0.058& ~0.011&28&0.96&1.1&0.15&0.17\\
3 &224  &0.16&0.067&$<$0.010&50&0.99&1.0&0.19&0.20\\
4A&222  &0.20&0.067&~0.020&26&0.77&0.91&0.23&0.24\\  
4B&227  &0.21&0.067&~0.026&15&0.52&0.85&0.24&0.26\\
5 &227&0.23  &0.081&~0.027&20&0.70&1.0&0.26&0.28\\
\hline
WM&     &0.80&0.29 &0.098 &  &4.3 &   &0.90&1.0\\
\hline
\end{tabular}
\end{minipage}
\end{table*}

The best three-component fits to each Epoch are listed in
Table~\ref{tab:sl2fits} and plotted with the observations and continua
in Figure~\ref{fig:sl2}. Derived optical depth profiles are plotted
with the modelled laboratory components in
Figure~\ref{fig:optplot}. The increases in $c_{pah}$, $c_{mel}$ and
$c_{SiC}$ with time will be discussed in Section~\ref{sec:67micvar}.

At Epochs~1 and~3 the observational uncertainties are too large giving
confidence intervals $>50$ percent and little improvement in
$\chi^2_\nu$ with three components and the SiC contribution is quoted
as an upper limit. Fits to Epochs~2, 4A, 4B and 5 were slightly better
with a bSiC component (improvement in $\chi^2_\nu$ 0.05--0.33) and the
1-sigma confidence intervals for bSiC were $\sigma_{bSiC} <30$ per
cent, with little change in $c_{pah}$ (0-10 per cent lower) and
$c_{mel}$ (increased by 7--10 per cent) for all but Epoch~5, where
$c_{pah}$ and $c_{mel}$ decreased by 50 per cent and 4 per cent,
respectively. The improvement in fit quality was due to the 6.2~ and
6.6~\micron\ bands combining with the 6.3-\micron~PAH band to broaden
it and flatten the interval between it and the 6.9~\micron\ band
(compare solid and dot-dash curves in Figure~\ref{fig:optplot}). The
modelled bSiC components will be extrapolated in
Section~\ref{sec:abspec} to show the influence of weaker 7.1 and
7.6~\micron~overtones.

\begin{figure}
\includegraphics[bb=125 205 452 590,width=\linewidth,clip=]{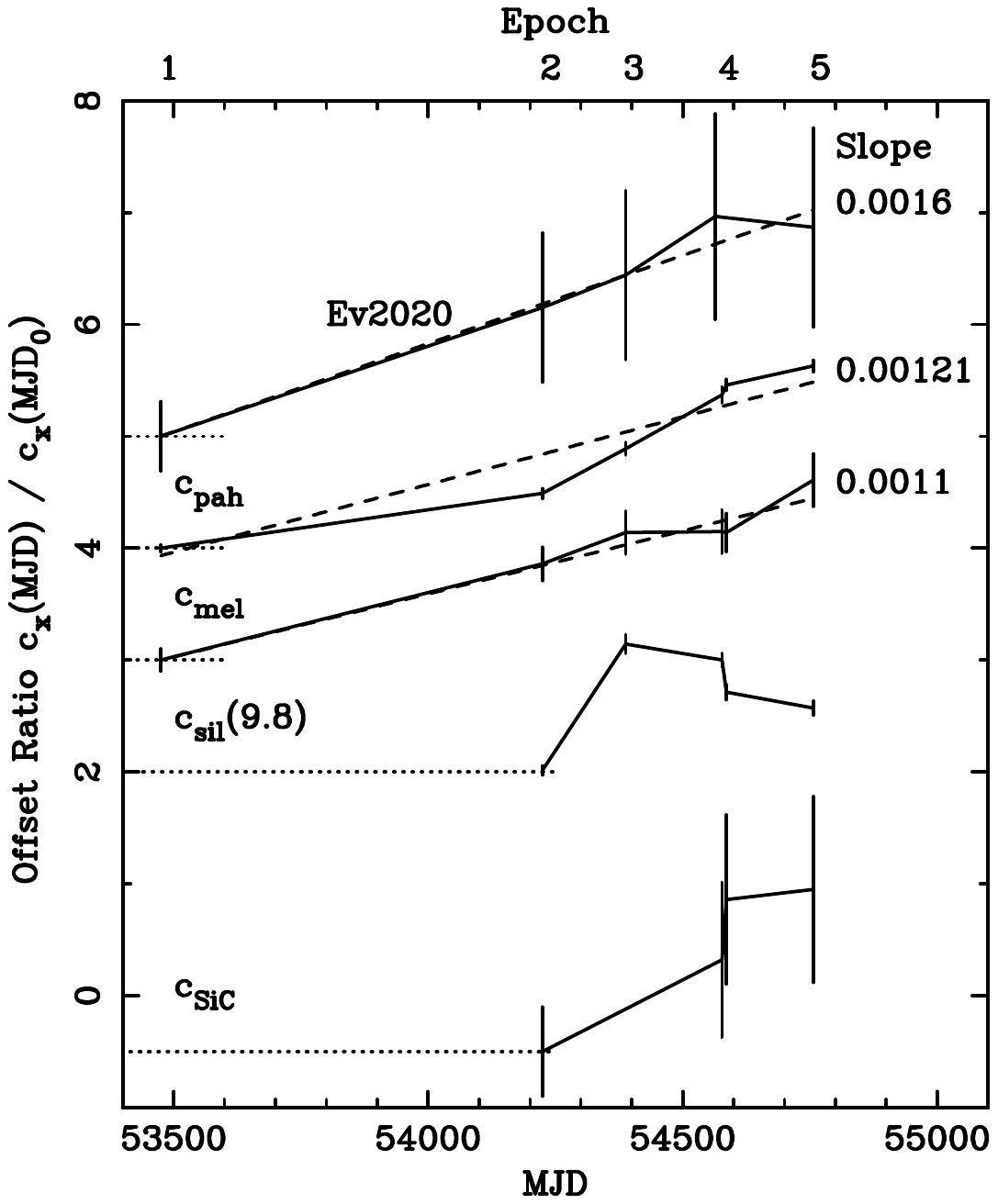}
\caption{Increases in dust mass during the Spitzer observing
  period. Values obtained by \citetalias{Evans2020} are from the flux
  method. $c_{pah}$, $c_{mel}$ and $c_{SiC}$ are the fitted optical
  depths of the small PAH and melilite and bSiC overtones,
  respectively. Each indicator, except $c_{SiC}$, is ratioed to the
  value at Epoch~1 (2005-04); $c_{SiC}$ is ratioed its value at
  Epoch~2 (2007-05).  Offsets are indicated by dotted lines. Slopes in
  the figure are in optical depth per day, Uncertainties in the fitted
  slopes are 0.0005 \citepalias{Ev2020}, 0.00004 ($c_{pah}$), 0.0001
  ($c_{mel}$) per day.\label{fig:dmi}}

\end{figure}

\subsection{Increases in PAH, melilite and bSiC with time}
\label{sec:67micvar}

The fitted PAH, melilite and bSiC optical depths at each Epoch are
listed in Table~\ref{tab:sl2fits} and plotted against their modified
julian dates (MJD; Table~\ref{tab:observations})
Figure~\ref{fig:dmi}. In the Figure each parameter has been ratioed to
its value at the earliest Epoch, MJD$_0$, with a proper detection (not
an upper limit). Hence, $c_{pah}$ and $c_{mel}$ are ratioed to values
at Epoch~1, whilst $c_{SiC}$ is ratioed at Epoch~2.

Slopes of linear fits to $c_{pah}$ and $c_{mel}$ indicate increase
rates of $0.44\pm 0.01$ ~yr$^{-1}$ and $0.40\pm 0.04$
~yr$^{-1}$. These rates are consistent with \citetalias{Evans2020}'s
$0.6 \pm 0.2$~yr$^{-1}$ estimate of the rate of increase in dust mass
from the emitted flux ($\lambda F_\lambda$) within the uncertainties,
even though the absorbing grains may have been formed at an earlier
time. Within the 1-sigma confidence intervals, the slope of $c_{pah}$
appears to increase to $0.99 \pm 0.06$ yr$^{-1}$ during Epochs 2--4B
before readopting the slower rate of increase between Epochs~4B
and~5. $c_{mel}$ increases between Epochs~1 and 5. There is an
apparent increase in bSiC with a jump between~4A and~4B but the
uncertainty in this is large.

\subsection{Weighted mean optical depth profile}
\label{sec:wmean}
\begin{figure}
\includegraphics[bb=125 160 440 400,width=\linewidth,clip=]{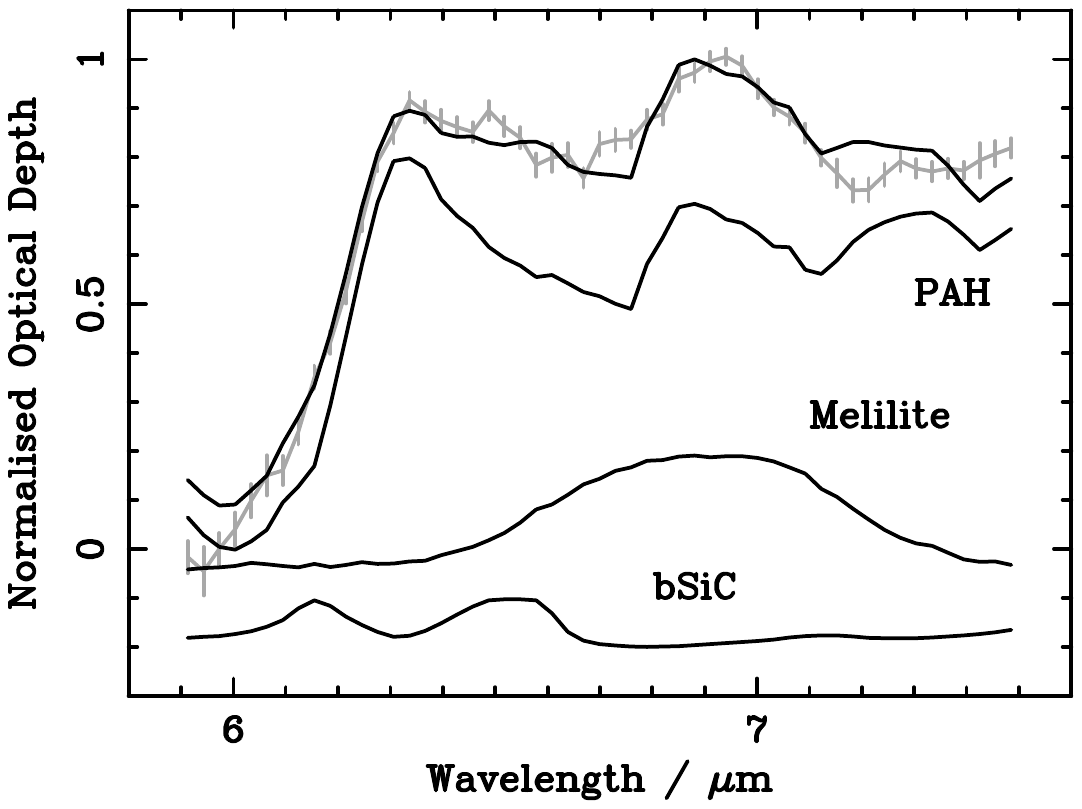}
\caption{Weighted mean of normalised optical depths (grey) with 3 component fit (black) and components below. \label{fig:moptplot}}
\end{figure}

Since variation in the relative proportions of the dust components is
small it is possible to obtain a higher signal-to-noise time-averaged
6--7~\micron\ dust profile for comparison with dust features other
environments such as molecular clouds and YSOs. The profile was
obtained after normalising the profiles from Epochs~1 to 5 at
6.9~\micron.  The weighted mean dust profile, together with fits of
the PAH, melilite and bSiC components, is presented in
Figure~\ref{fig:moptplot}. As expected, fits to the PAH, melilite
and bSiC features resemble those of the strongest features observed in
Epoch~5; when scaled to $\uptau_{6.9}$ they are, 0.22, 0.81, 0.027,
respectively: only $c_{pah}$ is slightly lower, reflecting its faster
growth rate.

\section{Fitting 8.4--13.3~\micron~spectra}
\label{sec:sl1mod}
\begin{figure*}
\includegraphics[bb=0 320 266 755,width=0.7\linewidth,clip=]{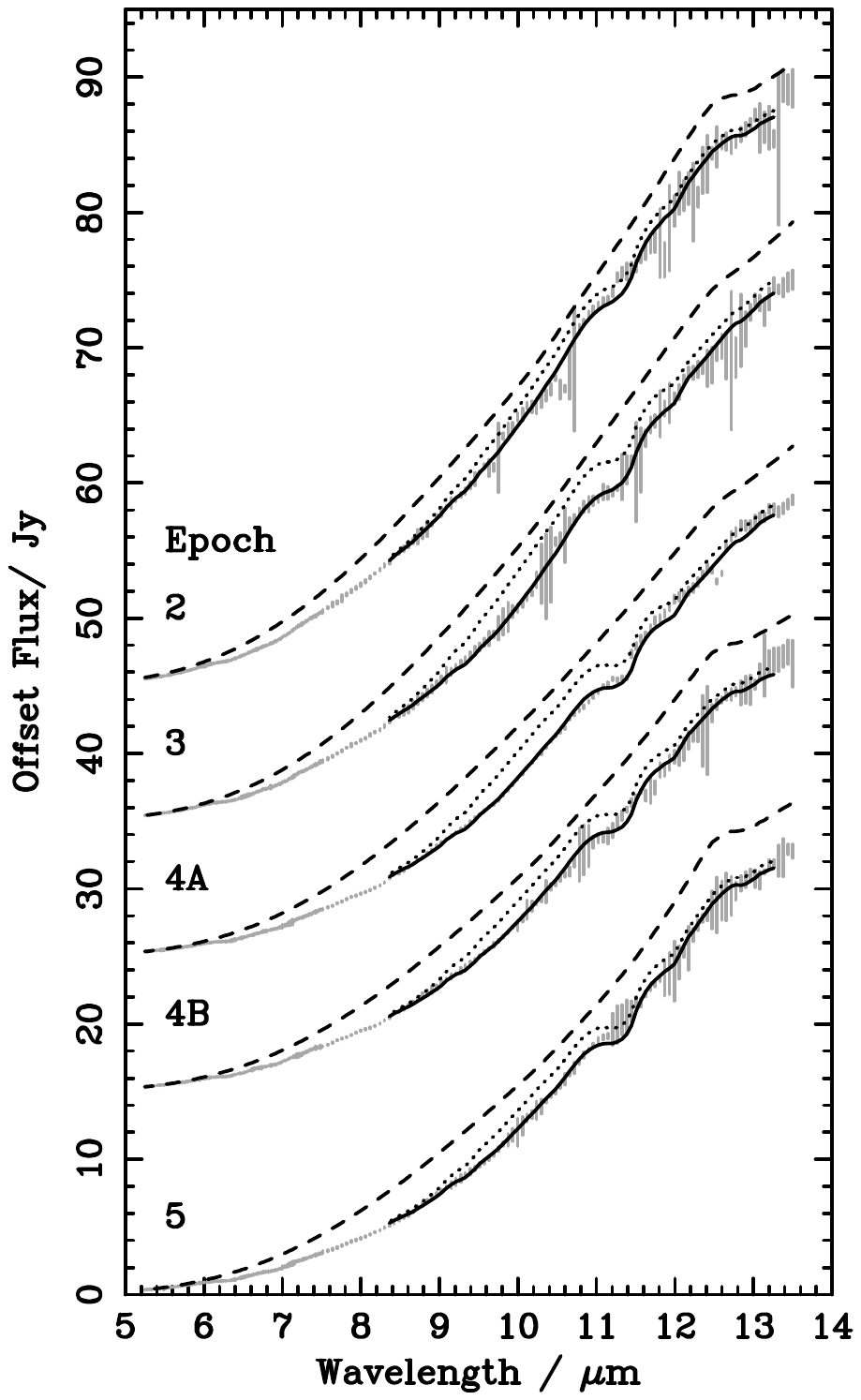}
\caption{Fits between 8.3 and 13.3~\micron. (a) 8.3 and
  13.3\micron\ fits (solid curves) to the SL1 flux spectra (grey error
  bars) using the PAH fit to the SL~2 data; dotted curves -- effect of
  removing foreground silicate absorption; dashed curves - modelled
  continua when the foreground absorption is zero. Y-axis offsets
  (bottom to top) are 0.0, 15, 25, 35 and 45 Jy.\label{fig:sl1fluxemiss}}
\end{figure*}

\begin{figure*}
\includegraphics[bb=0 280 520 478,width=\linewidth,clip=]{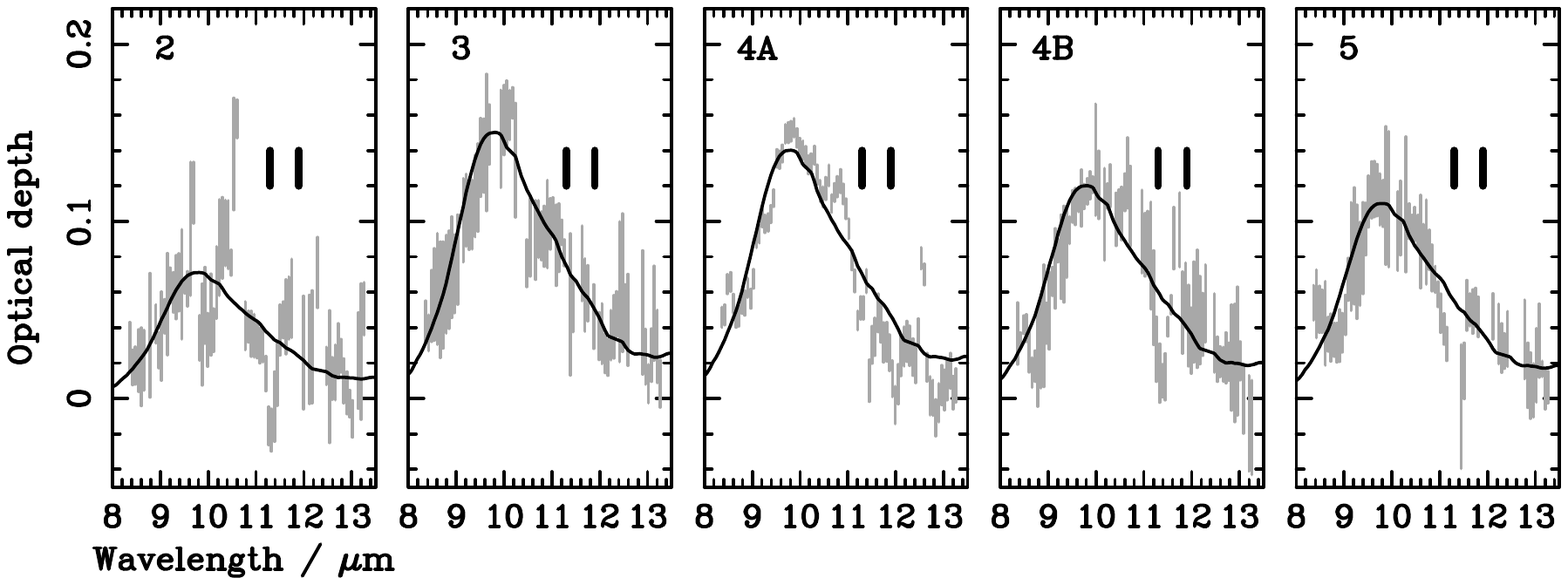}
\caption{ Changes in the 10-\micron\ astrosilicate
  absorption \label{fig:fitsilabs} compared with Cyg OB2 no. 12 after
  subtraction of the PAH component. Noisy points (uncertainty
  > 0.04 ) have been removed for clarity. Bars indicate the centres of PAH
  bands, which may be too deep.}.
\end{figure*}

\begin{figure}
\includegraphics[bb=0 496 270 674,width=\linewidth,clip=]{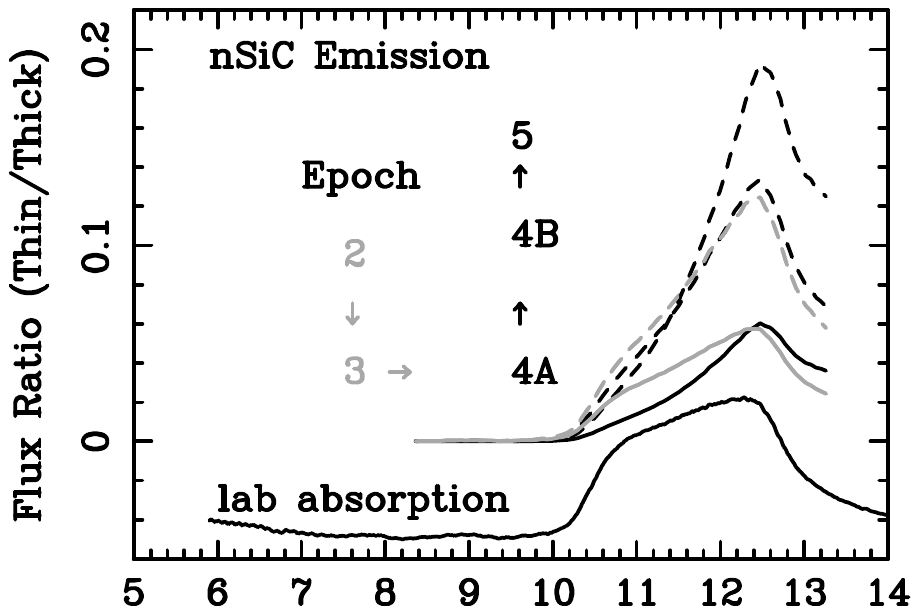}
\caption{Modelled nSiC emission: when the 11.3-\micron~PAH absorption band
matches the observations (solid), dashed when it does not. Grey and
black curves indicate falling and rising optically-thin fractions,
respectively. The laboratory profile of nSiC is shown below (arbitrary
normalisation; offset -0.05 in the Y-axis). \label{fig:sl1fluxemiss2}}
\end{figure}

Longer-wavelength observations at Epochs~3 and~4A appeared to have a
narrow absorption feature at
11.3~\micron~(Figure~\ref{fig:sl1fluxemiss}(a)) so a single-component
PAH model was extended to derive a continuum for it. Melilite and bSiC
components were not included because these grains are opaque in this
wavelength range.  However, fits with Equation~\ref{eq1} did not match
the overall shape of the flux spectrum until a foreground
'interstellar' 10~\micron~silicate absorption feature was added,
represented by the diffuse-medium dust profile towards Cyg OB2 no. 12,
hereinafter called astrosilicate; the feature shape, shown in
Figure~\ref{fig:sl1abs}(b) $\uptau_{sil}(\lambda)$ is derived in
Appendix~\ref{app:cyg}.

Fits were further improved when a single-temperature optically-thin
emission component due to nanometer-sized SiC grains, called nSiC (see Figure~\ref{fig:sl1fluxemiss2}),  was added
so that the source flux, $F_{\nu*}$ is given by,
\begin{equation}
F_{\nu*}\propto B_\nu(T_1)+B_\nu(T_2)c_{nSiC}\uptau_{nSiC}(\lambda)
\end{equation}
where T$_1$ and T$_2$ are the temperatures of the optically-thick and
optically-thin emission components and $c_{nSiC}$ is a scaling
constant for normalised nSiC emission, $\uptau_{nSiC}(\lambda)$. This emitted
continuum was then extinguished by foreground astrosilicate dust to give the observed flux, F$_\nu$,
\begin{equation}
  F_\nu =c'_0F_{\nu*}\exp{(-c_{pah}\uptau_{pah}(\lambda)-c_{sil}\uptau_{sil}(\lambda))}
\end{equation}
where, scaling constant $c'_0$, T$_2$ and $c_{sil}$ were fitted by $\chi^2$-minimization and 
$c_{pah}$ was fixed to the value obtained from the
5.9--7.5~\micron~fits. Continua for merged SL~1--SL~3 5--13.5~\micron~
spectra were derived by setting $c_{pah}$ and $c_{sil}$, to zero and
extending the wavelength range.

\subsection{Constraints and exclusions}
  
No attempt was made to model the 7.5--8.3~\micron-range for several
reasons: (i) there are inherent incertainties in the LRS observations
because of merging between orders, (ii) laboratory spectra of PAHs
show a great deal of variability in the shape of the defect band,
(iii) silicate overtones merge with Si--O bands making results highly
dependent on the baseline subtraction in the laboratory data
\citep[see][]{BH2005}, (iv) the bSiC and melilite grains
responsible for overtone features are a separate population to the
submicron-sized and nanometer-sized grains responsible for features
between 8--13~\micron~and it is not clear how to merge the data in these
simple models.

Since the PAH bands surround the 10-\micron~silicate absorption band
(Figure~\ref{fig:sl1abs}(b)) $c_{pah}$ was set to zero to explore the
interplay between them. In these cases statistically better fits were
obtained, the temperature was $\sim 20$-K cooler, optically-thin
emission negligible and the foreground silicate absorption very
weak. These fits were disregarded because the fluxes of these continua
fell well-below the fluxes of the merged 5--8~\micron\ spectra
which is inconsistent with a central stellar energy source.

\section{8.3--13.3~\micron\ Fits}
\label{sec:8--13fit}
\begin{table}
\centering
\caption{Fits to 8--13~\micron~SL~1 spectra of Sakurai's Object, $T_1$ (pre-set) and
  $T_2$ (fitted) are the temperatures of the optically-thick and optically-thin
  components, respectively. Sakurai was not observed at 8--13~\micron~during Epoch~1.}
\label{tab:sl1fit}
\begin{minipage}{\linewidth}
\begin{tabular}{llllllll}
\hline
Epoch&T$_1$&T$_2$&$\epsilon_{12.5}$ &$F_{12.5}$\footnote{Weighted mean flux between 12.0 and 13.0~\micron; $\pm$0.1 Jy plus order match uncertainty}&$c_{sil}$($\sigma$\footnote{1 sigma confidence interval per cent.})&$\chi^2_\nu$  &PAH?\\
     &K     &K&&Jy&&&11.3\\
\hline
2 &237   &119  &0.123&40.3    & 0.071 (3)&  3.5	&--  \\
3 &225     &137  &0.057&33.8  & 0.15 (2)&  0.77&Y\\
4A&225     &81   &0.059&28.7  & 0.14 (0.7)&  8.4\footnote{Reflects only RMS errors; CASSIS did not output systematic errors for SL~1}	&Y   \\
4B&228   &100  &0.132&28.5 & 0.12 (2)&  1.8	&--  \\ 
5 &228   &75   &0.191&28.0 & 0.11 (2)&  1.2 &--   \\
\hline
\end{tabular}
\end{minipage}
\end{table}

The observations (grey error bars), fits (solid curves) and continua
(dashed curves) are displayed in Figure~\ref{fig:sl1fluxemiss}. If the
silicate absorption component is excluded the fits do not match the
curvature of the observations between 11 and 12~\micron\ (dotted
curves). Fitted parameters are listed in
Table~\ref{tab:sl1fit}. 11.3~\micron~absorption bands in Epochs~3 and
4A are well-matched by the modelled SL~1 PAH strengths, but the
absorption bands are not observed at other Epochs. Epoch~2 might have
a narrow PAH emission band at 11.3~\micron, but this may be
coincidental with the noise.

Fits to all Epochs required both optically-thick and optically-thin
emission components and the proportion of optically-thin nSiC emission
varied with time. The temperature of the optically-thick component was
usually 1--3-K warmer than the 6--7~$\mu$m continuum, but Epoch~2
required a 11~K higher blackbody temperature and weaker foreground
silicate absorption ($\uptau\sim 0.07$); the difference is caused by
narrow-band structures in the spectrum of unknown origin at 9.7~ and
10.5\micron\ -- either unidentified gas-phase absorption or noise
since the error bars are large. The temperature of the optically-thin
component, $T_2$, settled on values 50--155~K cooler than the
optically-thick emission but the absence of an identifiable trend
indicates it is not well-constrained.

\subsection{Temporal variation in the 10~\micron\ astrosilicate absorption}
Fitted optical depths of the foreground silicate absorption,
$c_{sil}$, are within $\pm 0.04$ of the $0.105\pm0.005$ value deduced
by \citet{Evans2002} and modelled values increase between Epochs ~2
and ~3 before decreasing between Epochs~3 and 5 (Figure ~\ref{fig:dmi}
and Table~\ref{tab:sl1fit}). The silicate absorption profiles are
shown in Figure~\ref{fig:fitsilabs} after subtraction of the modelled
PAH absorption: the feature in Epoch~2 is not really detected above
the noise but changes in the features between Epochs~3 and 5 are
significant and inconsistent with a homogeneous distribution of
\emph{interstellar} silicate along the line of sight which is
unrelated to Sakurai's Object.  It is plausible that some of
the astrosilicate absorption is due to dust distributed within the
ancient PN and that disturbances associated with the eruptive
dust-formation event are causing silicates to coagulate into
larger grains. The interstellar feature towards Cyg OB2 no.~12 dust
can still be used for modelling because the profile is similar in
shape to the emission from new dust surrounding the O-rich AGB star
$\mu$~Cephei \citep[e.g][]{Bowey2004}. The relationship between the
decrease in astrosilicate and possible increase in melilite in Epoch~5
will be explored in Section~\ref{sec:orichevo}

\subsection{Temporal variation of optically-thin nSiC emission}
\label{sec:sicemiss}
The ratios of the optically-thin emission to the optically-thick
emission at each Epoch are compared in
Figure~\ref{fig:sl1fluxemiss2}. The optically-thin component is
smallest at Epochs~3 and 4A, in which the 11.3-\micron~absorption
feature matches the 6--7-\micron~fits. The fraction of optically-thin
emission is largest when the fitted silicate absorption is weakest
indicating that during fitting the two parameters work together to
increase the curvature of the flux spectrum to make it rise at longer
wavelengths.

Both the the total flux and the proportion of optically-thin emission
decreased between Epochs 2 and 3. Between Epochs~3 and 4A the flux
decreased more slowly and the optically-thin component did not change
significantly (though it might have cooled). The flux did not
change between Epochs~4A and 5, but became significantly more
optically-thin, changing rapidly in the in the 9 days between
Epochs~4A and~4B (April~2008). Unfortunately it is impossible to know
if the apparent increase in the optically-thin component in the 171
days between Epochs~4B and~5 occurred as rapidly as the change in the
days preceeding it.

\subsection{5.9--13.3~\micron\ Absorption Components}
\label{sec:abs}
\begin{figure}
\includegraphics[bb=280 125 540 765,width=\linewidth,clip=]{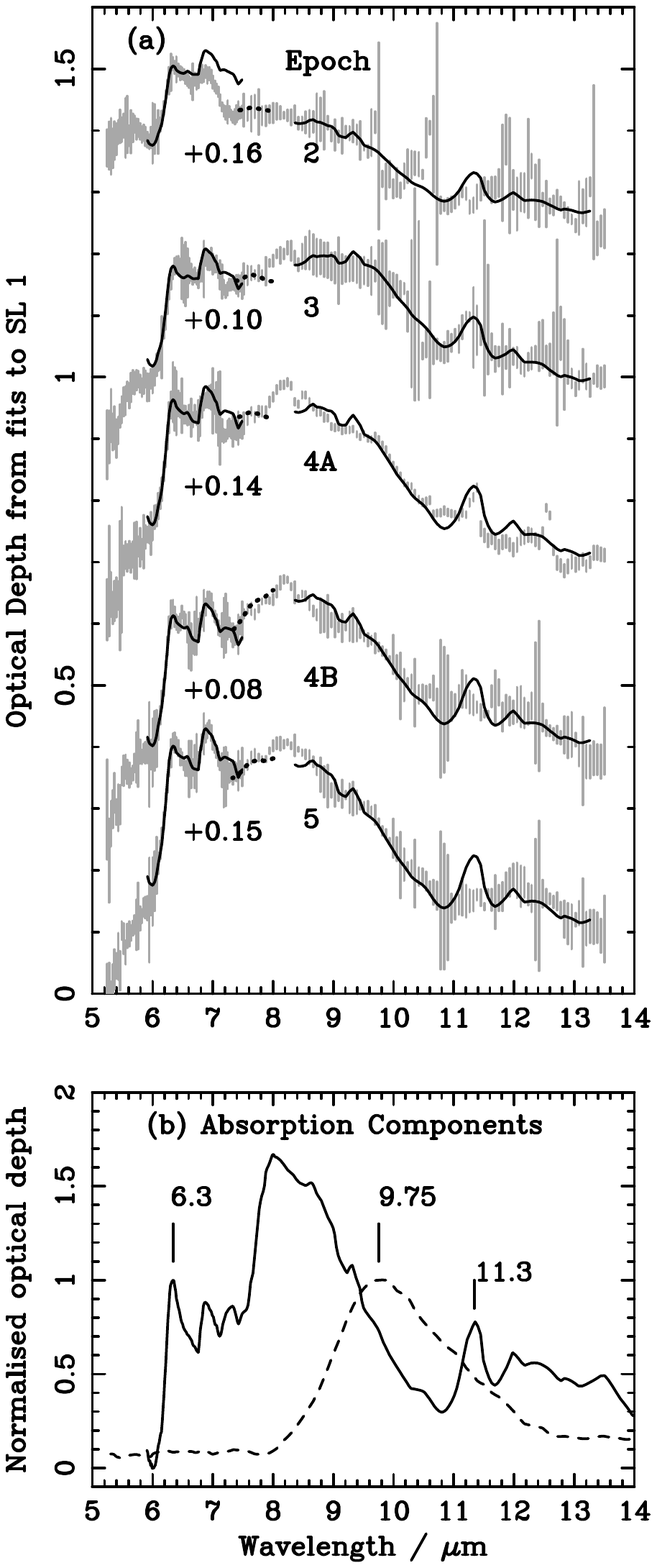}
\caption{(a) Optical depth profiles (grey error bars) obtained with
  modelled 8.3--13.3~\micron~continua; offsets in the Y-axis are 0.0,
  0.3, 0.6, 0.9 and 1.2; 8--13~\micron\ black solid curves are the
  fitted absorption features, 6-7.5~\micron\ are the fitted features
  from the SL~2 offset in the Y-axis to match the level of the
  longer-wavelength fits (additional offsets are indicated); dotted
  curves between 7.5 and 8.0~\micron\ indicate the contribution from
  bSiC extrapolated from the SL~2 fits. (b) Absorption profiles used
  to model 8.3--13.3~\micron\ spectra: PAH spectrum (solid) and
  silicate absorption (dashed) modelled with the profile of
  interstellar dust towards Cyg OB2 no. 12. Profiles are normalised at
  6.3~ and 9.75~\micron, respectively.\label{fig:sl1abs}}
\end{figure}
Optical depth profiles for Epochs~2--5 are compared with absorption
profiles fitted to the 8.3--13.3~\micron\ and 6--7.5~\micron\ data
(solid curves) in Figure~\ref{fig:sl1abs}(a); short dotted curves show
an extrapolation of the bSiC contribution
(Section~\ref{sec:bigsic}). Offsets are added to the
6--7.5~\micron\ fits to allow for the difference in scaling constants
$c_0$ and $c'_0$. The discrepancy between the slopes of the
6--7.5~\micron\ fit and the profile for Epoch~2 is caused by the 11~K
difference in the temperature of the optically-thick emission.

\subsection{PAH and bSiC}
\label{sec:abspec}
The 11.3-\micron\ PAH absorption features in Epochs~3 and 4A match the
band strengths implied by 6--7.5~\micron\ values of
$c_{pah}$. However, the observed 11.3~\micron\ PAH absorption is
negligible in Epochs~4B and 5. The feature might occur in emission at
Epoch~2. It is conceivable that the observed changes are consistent
with the trend in nSiC emission because as time progresses the PAH
feature moves from emission, to absorption followed by the emission
feature filling in the absorption feature - but there is insufficient
data to prove this.

There is also some correlation between the shape of excess absorption
near 8.0~\micron~and the shape and position of the C=C defect band
(Figure~\ref{fig:sl1abs}(b)) even though this is probably suppressed
by unconsidered emission or absorption components in the line of
sight. Surprisingly, the extrapolated weak bSiC features identified in
the 6--7.5\micron\ data resemble the shape of weak structure in the
merged spectra between 7.2~ and 8.0~\micron\ but the uncertainties are
large. If the features are real, the 6.0-- 8.0-\micron~spectral region
is the place to reconcile astronomical and meteoritic studies of SiC
grains. The existence of warmer nSiC grains and cool bSiC grains is
consistent with \citet{Speck2005}'s hypothesis that SiC grains become
smaller as the star evolves.

\section{Summary of observational results}
\label{sec:csil}

During the period of the Spitzer observations:
\begin{enumerate}
\item The mid-infrared flux decreased. The 5.6 and
  12.5~\micron\ fluxes decreased rapidly between Epochs 1 and 2, but
  the rate of change slowed between Epochs 2-4A and stopped between
  Epochs~4A and~5.
\item The proportion of optically-thin emission decreased between
  Epochs 2 and 3, did not vary between Epochs~3 and 4A (though it
  might have become cooler) before increasing rapidly in the 9 days
  between Epochs~4A and~4B and increasing more slowly into Epoch~5.

\item The mass of cold carbonaceous dust increased. PAH absorption
  increased throughout the observations, possibly with the rate
  accelerating during Epochs 2--4B, and then slowing between Epochs 4B
  and 5.  Concurrent increases in bSiC abundance might be significant,
  but are particularly affected by the low sensitivity to the overtone
  features at Epochs 1--3.

\item The apparent composition of the cold oxygen-rich dust changed
  with more melilite and less astrosilicate. The melilite absorption
  increased between Epochs~1 and ~5.  The astrosilicate abundance
  decreased between Epochs~3 and~5.

\end{enumerate}
  
Some of the spectral changes are subtle and their significance
affected by observational sensitivities. I believe changes in the
features to be real because they progressed between multiple
observations and some features were modelled independently in
different spectral ranges.

\section{Estimates of mass column density, number density and dust mass}
\label{sec:est}
\subsection{Mass Column Density and Number Density}
\label{sec:massdensity}

\begin{table*}
\centering
\caption{Column Mass Density and Number Density for Carbonaceous and Oxygen-rich dust Components}
\label{tab:codensity}
\begin{minipage}{\linewidth}
\begin{tabular}{lllllllllll}
\hline
&\multicolumn{6}{c}{Carbonaceous Dust}&\multicolumn{4}{c}{Oxygen-rich Dust}\\
Grain&\multicolumn{3}{l}{Mass Density}&\multicolumn{3}{l}{Number Density}&\multicolumn{2}{c}{Mass Density}&\multicolumn{2}{c}{Number Density}\\

Size&&&&53~nm&25~\micron&3~nm&&& 12~\micron&$<0.3$~\micron\\

Dust & PAH    & bSiC  &nSiC& PAH    & bSiC  &nSiC& melilite&astrosilicate& melilite&astrosilicate\\

&\multicolumn{3}{c}{$10^{-6}$~gcm$^{-2}$}&\multicolumn{3}{c}{cm$^{-2}$}&\multicolumn{2}{c}{$\times10^{-6}$~gcm$^{-2}$}&\multicolumn{2}{c}{cm$^{-2}$}\\

\hline
Epoch&&&&&&&&&&\\

1 &   $1.4_{0.71}^{6.5}$&&--      &2.4$\times 10^{10}$&   & --              &140&--   &   27000    & --               \\
\\
2 &   $2.2_{1.1}^{10}$&46 &0.082&3.8$\times 10^{10}$& 920      & 9.5$\times 10^{11}$    &260&27   &   50000    & 3.0$\times 10^{8}$\\
\\
3 &   $2.7_{1.3}^{12}$& &0.038&4.7$\times 10^{10}$&& 4.4$\times 10^{11}$     &310&58   &   60000    & 6.5$\times 10^{8}$\\
\\
4A &  $3.3_{1.7}^{15}$&83 &0.039&5.7$\times 10^{10}$& 1660     & 4.5$\times 10^{11}$    &310&54   &   60000    & 6.1$\times 10^{8}$\\
\\
4B &  $3.5_{1.7}^{16}$&110&0.088&6.0$\times 10^{10}$& 2200     & 10$\times 10^{11}$     &310&46   &   60000    & 5.2$\times 10^{8}$\\
\\
5 &   $3.7_{1.8}^{17}$&113&0.13 &6.4$\times 10^{10}$& 2300     & 15$\times 10^{11}$     &370&42   &   71000    & 4.7$\times 10^{8}$\\
\hline
WM\footnote{PAH, bSiC and melilite -from fits to weighted mean spectrum scaled to  mean of Epochs 3--5 ($\uptau_{6.9}=0.25$); nSiC and astrosilicate from mean of Epochs 3--5.} &$3.3_{1.7}^{15}$&100 &0.073&5.7$\times 10^{10}$& 2000  & 8.4$\times 10^{11}$&330&50&68000& 5.6$\times 10^{8}$\\
\hline
\end{tabular}
\end{minipage}
\end{table*}

An estimate of the mass column density of dust surrounding the region of
optically-thick emission can be obtained because the peak optical
depth, $\uptau_{pk}$, of an absorber is related to the mass absorption
coefficient, $\kappa_{pk}$:
\begin{equation}
  \uptau_{pk}=\rho \kappa_{pk} L,
\label{eq:taukap}
\end{equation}
where $\rho$ is the effective mass density of the absorber and $L$ is
the path-length of the light through the dust so that the mass column density of absorbers, $\Sigma=$ along the line of sight is given by:
\begin{equation}
  \Sigma=\rho L= \uptau_{pk}/\kappa_{pk}.
\end{equation}

The column number density of grains is obtained by dividing $\Sigma$,
by the single-grain masses, $m_g$, in Table~\ref{tab:lab}.  The
mass and grain column densities are listed in Table~\ref{tab:codensity}.

\subsection{Rates of increase in Mass Density and Number Density}
Equation~5 and the rates of change of PAH and melilite optical-depths
obtained in Section~\ref{sec:67micvar} can be used to deduce the rate
of increase in mass and number density: the average rate of increase
in PAH density between Epochs 1--5 is $d\Sigma_{pah}/dt=7.3 \times
10^{-6}$ gcm$^{-2}$yr$^{-1}$ ($1.3\times 10^{11}$ grains
cm$^{-2}$yr$^{-1}$) and the enhanced rate between Epochs 2 and 4B is
$d\Sigma_{pah}/dt=17\times 10^{-6}$ ($2.8 \times 10^{11}$ grains
cm$^{-2}$yr$^{-1}$) For melilite, the mean rate of mass density
increase, $d\Sigma_{mel}/dt =1.8 \times 10^{-3}$gcm$^{-2}$yr$^{-1}$ or
$0.35\times 10^6$ grains cm$^{-2}$yr$^{-1}$.

\subsection{Dust Mass estimates and their rate of increase}

Since mass column density is related to the effective mass density of
a dust grain, Equation~\ref{eq:taukap} can be used to estimate the
total dust mass of each component at each Epoch with the caveats that
neither the distance to Sakurai's object, nor the geometry of the
dusty region associated with it, nor the mass absorption coefficent of
PAHs are well-defined (see Appendix~\ref{app:pah} for my derivation of
$\kappa_{pk}$). Estimates of the distance to Sakarai's Object vary
considerably \citep[reviewed by][]{Hinkle2014}. I adopt a distance of
3.5kpc which is the maximum used by \citet{Chesneau2009} in models of
their mid-infrared observations and consistent with
\citetalias{Evans2020}'s preferred value of $3.8\pm0.6$kpc.
\citet{Chesneau2009} obtained mid-infrared interferometer observations
of Sakurai's object on 2007 June 29--30 (MJD~54280--54281; 55 days
after Epoch~2) and deduced that the source is surrounded by an opaque
dusty torus or thick disk of $105 \times 140$AU and a scale height of
$47\pm 7$AU.

Assuming that the cold PAHs and bSiC grains are located in the torus I
use a cylindrical geometry to give a volume of $\pi R^3$, where R is
50~AU and the optical path length from the optically-thick shell $L=R$
so that the mass, $M$, of dust in the torus due to an absorber is
approximated by
\begin{equation}
  M=\Sigma \pi R^2
\end{equation}

\begin{equation}
M_{50 AU}\approx 8.8\times 10^{-4}\Sigma \ M_{\sun}
\end{equation}
and the rate of mass increase can be deduced by replacing $\Sigma$
with $d\Sigma/dt$. An inferred mass of nSiC is not given because the optically-thin emission
probably derives from a region closer to the optically-thick
`photosphere' of unknown volume.

For the 50AU torus, $d\Sigma_{pah}/dt$ equate to an average mass
increase of 6.4$\times 10^{-9}$~M$_{\sun}$ yr$^{-1}$ and an
accelerated rate between Epochs~2 and~4B of 16$\times
10^{-9}$~M$_{\sun}$ yr$^{-1}$; the faster rate is similar to
\citet{Evans2020} estimate in May 1999, which was 11$\times
10^{-9}$~M$_{\sun}$ yr$^{-1}$, when scaled to the PAH density, but
lower than the rate of 40$\times 10^{-9}$~M$_{\sun}$ yr$^{-1}$
in Sept 2001.

\section{Evolution of the dust}
\label{sec:evo}

\subsection{PAH, bSiC and nSiC}

\begin{table*}
\centering
\caption{Estimated masses of carbonaceous dust  (PAH and bSiC) in the 50~AU torus
  compared with \citetalias{Evans2020} estimates by the peak flux ($\lambda F_{\lambda}$)
  method after conversion to PAH or SiC mass densities instead of amorphous carbon ($\rho=1.5 $gcm$^{-3}$) and scaled by 0.85 to reflect the choice of D=3.5~kpc.}
\label{tab:cmass}
\begin{minipage}{0.75\linewidth}
\begin{tabular}{lllrllllllll}
\hline
&&\multicolumn{2}{c}{6--7\micron\ mass}&$\lambda$F$_\lambda$& \multicolumn{4}{c}{$\lambda$F$_\lambda$ mass similar to 6--7\micron\ mass}\\
&& PAH    & bSiC  &PAH& & PAH &SiC & T\\
Ep.&Date\footnote{day-month-year}&\multicolumn{3}{c}{$10^{-9}$~M$_{\sun}$}&Date&\multicolumn{2}{c}{$10^{-9}$~M$_{\sun}$}&K\\
\hline
1  &15-04-2005&1.2  && 1100 & 18-08-1998&& 0.92  &1210\\
2  &04-05-2007&1.9 &41&2400&21-04-1999       &3.9&32    &841 \\
3  &15-10-2007&2.4 &&2900&&&\\
4A &21-04-2008&2.9 &74&3400\footnote{Measured on MJD~54564, 8 April 2008}&&&\\
4B &30-04-2008&3.1 &96&--& 03-05-1999&14&120 &723\\
5  &18-10-2008&3.3 &99&3400&08-06-1999&16&130 &717\\
   &          &    &  &    &14-06-1999&16&130 &717\\
   &          &    &  &    &06-09-1999&24&200 &661\\
\hline
\end{tabular}
\end{minipage}
\end{table*}

My dust mass estimates are compared with \citetalias{Evans2020} masses
deduced from optically-thick emission in
Table~\ref{tab:cmass}. \citetalias{Evans2020} estimates have been
scaled to PAH and SiC mass densities, as appropriate, and a distance
of 3.5~kpc. Since it is impossible to distinguish between dust types
in their data, I have converted their values for amorphous carbon to
PAH and SiC masses and quoted both for blackbody temperatures between
the PAH condensation temperature of 850~K \citep{Helling1996} and
650~K. I assume that the bulk of their 200--300~K dust is formed of
PAHs. The black-body temperature of 1210~K on 18 August 1998
(MJD~51043; \citet{Geballe2002}) was low enough for SiC condensation,
but too hot for PAH formation so I assume that this is SiC dust. It is
plausible that the inferred absorbing 20~\micron\ bSiC grains were
formed at this stage followed by the nanometre-sized grains in the
modelled optically-thin emission. This interpretation is consistent
with \citeauthor{Speck2005}'s observational and laboratory-based
hypothesis that SiC grains in AGB stars grow to large sizes in early
condensation phases when the mass-loss rate is low but that they
become nm-sized when the stellar mass-loss rate increases.

My inferred masses of bSiC and PAHs for Epochs~1--5 are, respectively,
2 and 3 orders of magnitude lower than \citetalias{Evans2020}
estimates from the optically-thick emission on the same dates but our
methods measure different grain populations: \citetalias{Evans2020}
measurement of optically-thick emission infers the dust mass of the
coolest external edge of the expanding shell\footnote{The mass
  estimate is sensitive to increases in surface area, but insensitive
  to warmer conditions inside the shell where most of the condensation
  occurs. Therefore foreground cool dust includes SiC dust
  condensed at an early time, but this contributes a decreasing
  fraction of the total mass. Due to the increasing surface area of
  the expanding dust shell and the low temperature of the optically
  thick leading edge inferred dust masses at the 200--300~K
  temperatures will probably reflect some geometric effects but an
  effect of 2--3 magnitudes seems unfeasibly large.}. My method
estimates the mass of absorbing dust in front of the optically-thick
region. This dust is not emitting at 6--7~\micron\ so it must be
cooler; I agree, with \citetalias{Evans2020}, that this cooler dust
must be older than dust in the optically-thick shell.

PAH formation probably occurred from around 21 April 1999 (MJD~51289)
when the temperature was 841~K and the SiC dust mass was
32$\times10^{-9}$~M$_{\sun}$), a value similar to the
41$\times10^{-9}$~M$_{\sun}$ estimate of mass in large SiC grains in
Spitzer Epoch~2 (4 May 2007). The effect of this rapid condensation is
seen in \citetalias{Evans2020}'s third measurement on 3 May 1999 (MJD
51301) in which there is a nearly four-fold increase in mass and the
temperature has dropped by 100~K in the 12 days following the previous
measurement followed by little change both values in measurements 36
and 42 days later (08 and 14 June 1999).

The increases in the 6--7~\micron\ PAH absorption seem to record this
condensation event in the enhanced rate of change in the
6.3~\micron\ PAH absorption between Epochs 2 and 4B and the reduced
rate to Epoch 5.  The cold dust masses observed in the Spitzer data
during Epochs~2 and 4B (04 May 2007 and 30 April 2008) correspond to
the warm dust masses inferred by \citetalias{Evans2020} on 21 April to
03 May 1999. The slight change between Epochs~4B and ~5 might match
the small increase between 03 May and 08 June 1999. By September 1999
\citetalias{Evans2020} dust mass is 50\% higher than the June values,
but there are no more Spitzer observations with which to compare
it. If the PAH absorption continued to grow at the rate of 0.44
yr$^{-1}$ the 6.3\micron\ PAH absorption feature would have been
opaque ($\uptau_{6.3}\approx 4$) by July 2016 when \citet{Evans2020}
Sofia data were obtained and its non-detection is a consequence of
high opacity, rather than noisier data.

\subsection{Silicate coagulation in Epochs~3 to~5}
\label{sec:orichevo}

The decreases in the astrosilicate absorption and consequent increases
in the large grains of melilite might be a consequence of the
coagulation of smaller astrosilicate dust into larger grains within
the torus or PN. However, astrosilicate is assumed to be glassy whilst
the large melilites are crystalline and a mechanism would be required
to introduce crystallinity on a short timescale unless the
astrosilicate already included a varied mixture of crystalline
silicates with narrow bands that merge to form the assumed smooth
10~\micron\ absorption feauture \citep[see][]{BA2002}. This form of
silicate dust is far more consistent with meteoritic and terrestrial
materials than entirely amorphous dust.

Classic astrosilicates are also Mg-rich with between olivine and
pyroxene stoichiometry $\sim$Mg$_{1.5}$SiO$_{3.5}$
\citep[e.g.][]{Fogerty:2016} with 1.5 Mg atoms per Si atom, whilst the
approximate stoichiometry of Si-rich melilite is
Ca$_{1.5}$Mg$_{0.5}$SiO$_{3.5}$, i.e. 1.5 Ca atoms per Si atom and 3
Ca atoms for every Mg atom. In the diffuse medium the ratio of Ca to
Mg atoms depleted from the gas into the dust is 8.4 per cent
\citep{SW1996}. Hence, if the available Ca were entirely located in
melilites up to 2.8 per cent of the silicate molecules or 2.5 per cent
of the astrosilicate mass could have had melilite stoichiometry. This
upper limit is similar to the 1.5--3 area per cent abundance of CAIs in
chondritic meteorites and more than 80 per cent of the CAIs are formed
of melilite or its alteration products (Section~\ref{sec:omel}). It is
also plausible that the melilites are the only coagulated components
observed due to the strength of its overtone band which is 3 times
that of other anhydrous crystalline silicates. I conclude that the
astrosilicates may have clumped together to form the inferred
melilites observed in the overtone features and that it is plausible
that grains like these were precursors to the melilite in the CAIs
in meteorites.

\section{Conclusions}
\label{sec:conclusion}
The mid-infrared flux decreased during the period of the Spitzer
observations. The 5.6 and 12.5~\micron\ fluxes decreased rapidly
between April 2005 and May 2007, but the rate of change slowed between
May 2007 and April 2008 and stopped between April and October 2008.

The proportion of optically-thin emission due to nanometre-sized SiC
grains decreased between May 2007 and an apparent minumum in October
2007 which continued until 21 April 2008. The optically-thin
proportion then increased rapidly in the 9 days to 30 April 2008 and
then more slowly in the six months to October 2008. 

The mass of cold carbonaceous dust increased. PAH absorption increased
throughout the observations, with the rate accelerating between May
2007 and May 2008 and slowing down in the six months to October 2008.
Increases in the abundance of $\la 25$\micron-sized SiC grains in
April 2008 might be significant but the data are not very sensitive to
the overtone features before this time.

SiC formation mainly occurred before May 1999, followed by PAH
formation in April--June 1999 once the dust was sufficiently
cool. These grains are likely to be responsible for the growth in
absorption due to PAH and $\la 25$~\micron-sized SiC dust between May
2007 and April 2008. The formation of large grains during the earlier
period of mass loss and later optically-thin emission from 3nm-sized
SiC is consistent with the hypothesis of \citet{Speck2005} that SiC
grains are large in early phases when the mass-loss rate is low but
that they become nm-sized when the infrared emission is optically
thick and the stellar mass-loss rate is high.

Much of the dust responsible for the weak 10~\micron\ silicate
absorption towards Sakarai's Object is likely to be associated with
Sakurai's PN rather than interstellar in origin, because the
optical-depth of the feature decreased between October 2007 and
October 2008. Decreases in the astrosilicate absorption and increases
in the large grains of melilite might be a consequence of the
coagulation of smaller astrosilicate dust into larger grains within
the torus or PN. Spectroscopic and Ca-abundance requirements are
satisfied if up to 3 per cent of astrosilicate is converted to grains
which produce melilite overtone features. It is plausible that grains
like these were the precursors to the melilite in the CAIs in
meteorites because more than 80 per cent of the CAIs are formed of
melilite or its alteration products and the area per cent abundance of
CAIs in chondritic meteorites is 1.5--3 per cent.

This study illustrates the importance of continued monitoring of
mass-losing objects similar to Sakurai's Object in the
mid-infrared. Intriguing results from this study are the appearance,
growth and possible and disappearance of the 6--7~\micron\ absorption
features and their use for the detection of PAHs and inference of 10~
to 20~micron-sized melilite and SiC grains which are otherwise
indetectable due to their high opacity in the 10~\micron\ Si--O and
Si-C stretching bands. The detection of melilite overtone absorption
bands in circumstellar dust around oxygen-rich stars in the absence of
H$_2$O ice and carbonaceous dust would be a priority, but it may be
that the grains in these environments are too small and the
grain-density too low to produce overtone features.

\section*{Acknowledgements}

The author would like to thank the anonymous referee for insightful
and careful reviews which have improved this paper. The author is
supported by a 2-year \emph{Science and Technology Research Council}
Ernest Rutherford Returner Fellowship (ST/S004106/1). Spectra of SiC
were supplied by A. M. Hofmeister.  LR Spitzer spectra were obtained
from the Combined Atlas of Sources with Spitzer/IRS Spectra (CASSIS),
a product of the \ Infrared Science Center at Cornell University,
supported by NASA and JPL.  This work is based on observations made
with the Spitzer Space Telescope, which is operated by the Jet
Propulsion Laboratory, California Institute of Technology under a
contract with NASA.

\section*{Data Statement}
Data generated in this article will be shared on reasonable request
to the corresponding author.

\appendix
\section{Laboratory and astronomical absorption profiles}
\subsection{Hydrogenated Amorphous Carbon (HAC)}
\label{app:HAC}

\begin{figure}
\includegraphics[bb=125 160 440
  400,width=\linewidth,clip=]{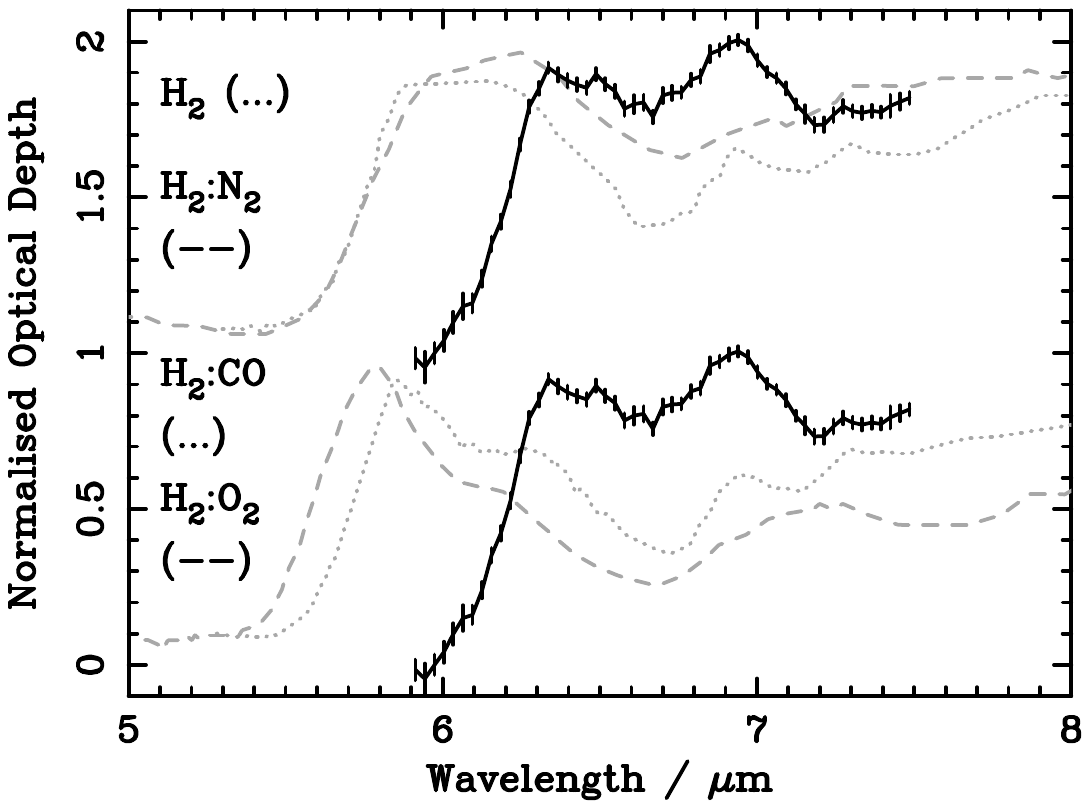}
\caption{Comparison between digitized spectra of HAC samples (grey, arbitrary normalisation)
  synthesized by from graphite in various gas mixtures
  \citet{Grishko2002} and the mean profile of Sakurai's Object (black)
  calculated in Section~\ref{sec:wmean}. The ratios of mixed gases
  during formation were 9:1. Fine structure below the resolution of
  the Spitzer spectra has been excluded.\label{fig:hacsak}}
\end{figure}

The mean optical depth profile of Sakurai's Object calculated in
Section~\ref{sec:wmean} is compared with various HAC samples
synthesized by \citet{Grishko2002} in Figure~\ref{fig:hacsak}.  HAC
fails to match the 6.3~\micron\ peak of Sakurai's Object due to excess
absorption at 5.8 to 6.0~\micron.

\subsection{Polycyclic Aromatic Hydrocarbons (PAHs)}
\label{app:pah}
Three spectra of well-characterised soot samples obtained by
\citet{Carpentier2012} were compared with data for Sakurai's
Object. Soot sample 3 (which is dominated by PAH units with many
defects and twisted rings) provided the best match to observed
6--7~\micron~bands\footnote{\citeauthor{Carpentier2012} found that the positions of
these peaks matched the positions of class C aromatic infrared bands
(AIBs) defined by by Peeters et al. 2002 from emission bands observed
in post-AGB objects (IRAS 13416 and CRL 2688), better than their other
samples which had fewer defects.}.
 The laboratory sample consisted of agglomerated
particles with diameters of $\sim30$~nm. I removed narrow bands in the
5--8-\micron~region due to H$_2$O vapour contamination in the
spectrometer and a broader carbonyl -C=O band at 5.821~\micron~due to
the use of oxygen in during sample preparation to produce the spectrum
presented here.

Due to the inhomogeneous nature and undetermined particle thickness of
the samples used for spectroscopy, \citeauthor{Carpentier2012} wisely
presented spectra with arbitrary optical depth scales. However,
astrophysical studies require a an estimate of the band strength using
equation~\ref{eq:taukap}, $\kappa_{pk}=\uptau_{pk}\rho L$, where
$\rho$ is the effective mass-density and $L$ is the optical path
length. \citet{Gavilan2017} used the same sooting apparatus to produce
films with peaks near 6.3~\micron\ with an optical depths $\uptau \sim
0.07$ so I shall adopt this value.  In a compressed sample $\rho$ will
be close to the density of the bulk solid. However, the soot particles
are uncompressed and the effective mass is highly dependent on the
degree of agglomeration and likely to be far less than that for a
`solid lump' of PAH ($\sim$ 1.8~gcm$^{-3}$). \citep{Rissler2013}
ascertained effective mass densities of soot of geometric diameters of
53-70~nm produced by a propane flame of 0.39--0.20~gcm$^{-3}$ based on
a primary particle density of 1.8~gcm$^{-3}$. Hence, the value of
$\kappa_{pk}$ could be $1.3\times10^4$~cm$^2$g$^{-1}$($\rho \approx 1.8
$gcm$^{-3}$), $6\times10^4$~cm$^2$g$^{-1}$ ($\rho \approx
0.39$~gcm$^{-3}$) or $12\times10^4$~cm$^2$g$^{-1}$ ($\rho \approx
0.2$~gcm$^{-3}$). I adopt the primary particle size of 0.53~nm and
density of 0.39~gcm$^{-3}$ preferred by \citet{Rissler2013} giving
$\kappa_{pk}$=$6\times10^4$~cm$^2$g$^{-1}$.

\subsection{Melilite, Hibonite and SiC}
\label{app:mhs}
\begin{figure}
\includegraphics[bb=115 431 445 760,width=\linewidth,clip=]{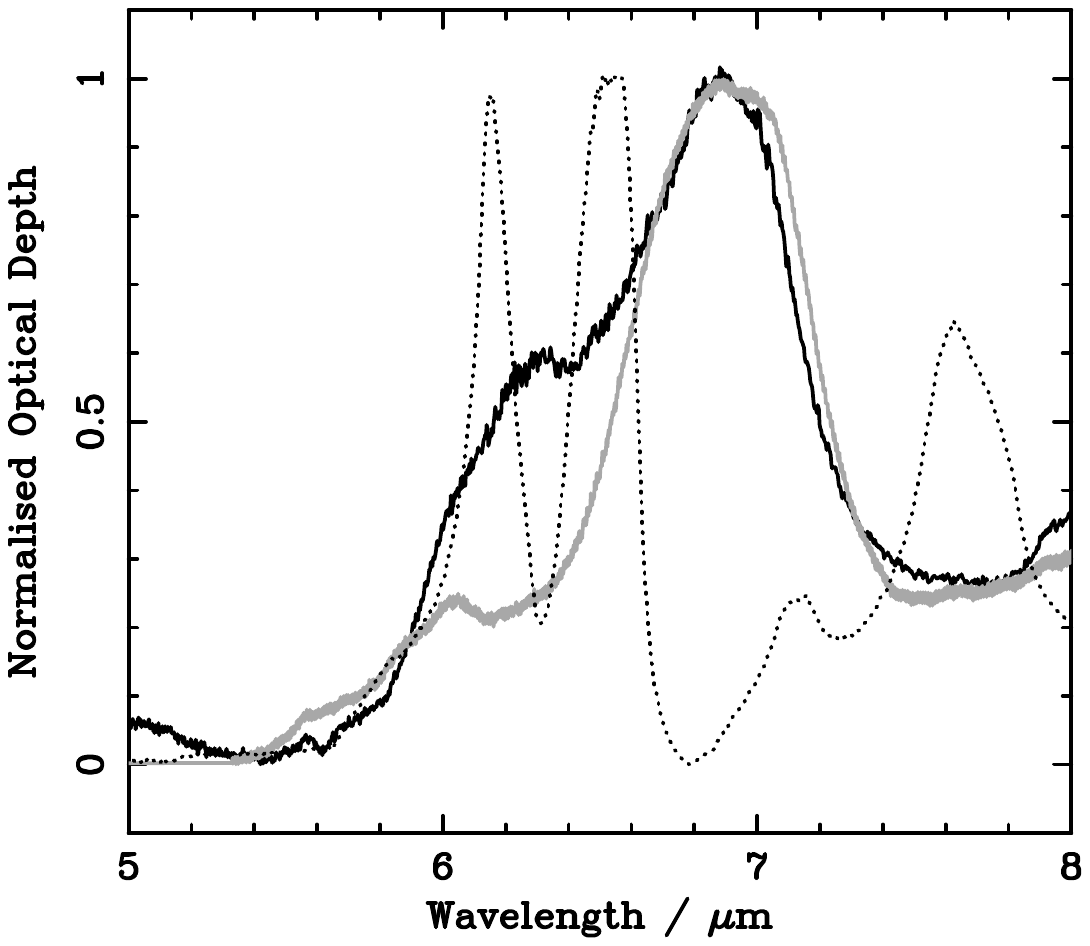}
\caption{Comparison between melilite(grey), metamict hibonite (black) and $\beta$~SiC overtones (dotted). Each spectrum is normalised to its largest peak. \label{fig:melhib}}.
\end{figure}

Normalised overtone spectra of compressed crystalline melilite and
metamict hibonite powders \citep{BH2005} are compared with the
normalised spectrum of a $\sim 25$~\micron-thick wafer of $\beta$~SiC
\citep{Hof2009} in Figure~\ref{fig:melhib}.

The spectrum of a compressed powder is representative of the feature
provided by a crystal of similar size because there are virtually no
spaces between the grains, but orientational effects are lost; this is
probably near to a lower limit for the melilite grain size because
powder was added until the overtone was observed. The metamict
hibonite overtone is broader than the melilite overtone due to a
subpeak at 6.3\micron. The band strength, $\kappa_{pk}(6.9)$ of
metamict hibonite is $0.48\times 10^2$cm$^2$g$^{-1}$. Other hibonite
overtones peak at 7.3~\micron\ and have even weaker $\kappa_{pk}$ of
0.08--0.14 $\times 10^2$cm$^2$g$^{-1}$ and do not have any peaks at
6.9~\micron.

The 25\micron\ thickness of the SiC wafer is likely to be an upper
size limit because larger grains are opaque, even in the overtones;
see \citet{Hof2009} for the spectrum of a 308~\micron\ sample. Ideally
a 20\micron\ wafer would have been measured because the
6.6~\micron\ peak is slightly rounded, but this risked destruction of
the specimen (Hofmeister, personal communication). Rounding of the
peak and other artefacts occur in specimens which are opaque at the
peak wavelength \citep[see][]{HKS2003}. The nano-$\beta$~SiC
spectrum \citep{Speck2005} is of a thin film of 97\% purity consisting
of 3~nm particles\footnote{This is their tabulated value; in the text
  it says the grains were 20~nm; values can be scaled accordingly}, of
thickness 0.34\micron\ \citep{Hof2009}.

\subsection{Cyg OB2 no. 12}
\label{app:cyg}
The shape of the interstellar silicate absorption feature in
Figure~\ref{fig:sl1abs}(b) was derived from a CASSIS Spitzer LRS
spectrum (AOR 27570176) calibrated by the method described in
Section~\ref{sec:obs}. Since the 8--13-\micron~spectrum was very
similar to ground-based observations obtained by \citet{Bowey1998} and
\citet{Bowey2004}, the continuum was estimated by scaling and
extrapolating a continuum derived by \citet{Bowey2004} to the
wavelength-range of the Spitzer SL~1 spectrum. The resulting
absorption profile was then normalised to unity at the peak wavelength
(9.75~\micron) and narrow lines (probably due to hydrogen emission
local to the star) removed before comparison with Sakurai data.  A
detailed and different analysis of silicate dust in this line of sight
has been published by \citet{Fogerty:2016}.

%%%%%%%%%%%%%%%%%%%%%%%%%%%%%%%%%%%%%%%%%%%%%%%%%%



%\bibliographystyle{mnras}


%%%%%%%%%%%%%%%%%%%%%%%%%%%%%%%%%%%%%%%%%%%%%%%%%%

%%%%%%%%%%%%%%%%%%%%%%%%%%%%%%%%%%%%%%%%%%%%%%%%%%

% Don't change these lines
\bsp	% typesetting comment
\label{lastpage}
\end{document}